\documentclass[a4paper]{jpconf}
\usepackage{graphicx}
\usepackage{amssymb}

\begin{document}
\title{Competition of local-moment ferromagnetism and superconductivity in Co-substituted EuFe$_2$As$_2$}

\author{M.\,Nicklas, M.\,Kumar$\footnote[1]{Present address: Institute for Solid State Research, IFW Dresden, 01171 Dresden, Germany}$, E.\,Lengyel, W.\,Schnelle and A.\,Leithe-Jasper}
\address{Max\,Planck\,Institute\,for\,Chemical\,Physics\,of\,Solids,
N\"{o}thnitzerstr.\,40, 01187\,Dresden, Germany} \ead{nicklas@cpfs.mpg.de (Michael Nicklas)}

\begin{abstract}
In contrast to SrFe$_2$As$_2$, where only the iron possesses a magnetic moment, in EuFe$_2$As$_2$ an
additional large, local magnetic moment is carried by Eu$^{2+}$. Like SrFe$_2$As$_2$, EuFe$_2$As$_2$
exhibits a spin-density wave transition at high temperatures, but in addition, the magnetic moments
of the Eu$^{2+}$\,order at around 20 K. The interplay of pressure-induced superconductivity and the
Eu$^{2+}$\,order leads to a behavior which is reminiscent of re-entrant superconductivity as it was
observed, for example, in the ternary Chevrel phases or in the rare-earth nickel borocarbides. Here,
we study the delicate interplay of the ordering of the Eu$^{2+}$ moments and superconductivity in
EuFe$_{1.9}$Co$_{0.1}$As$_2$, where application of external pressure makes it possible to sensitively
tune the ratio of the magnetic ($T_{\rm C}$) and the superconducting ($T_{\rm c,onset}$) critical
temperatures. We find that superconductivity disappears once $T_{\rm C}>T_{\rm c,onset}$.
\end{abstract}

\section{Introduction}
The discovery of high-temperature superconductivity in the iron-based superconductors has stimulated
an enormous interest in the study of this new class of materials. One peculiar finding is the
interplay of the local 4{\it f} moments of the Eu$^{2+}$ ions and superconductivity in EuFe$_2$As$_2$
under pressure, which is reminiscent of re-entrant superconductivity \cite{Miclea09,Terashima09}.
Like the (\textit{A}=Ca, Sr, Ba) members of the \textit{A}Fe$_2$As$_2$ family, EuFe$_2$As$_2$
exhibits a spin-density wave (SDW) transition around $T_0=190$~K related to the Fe$_2$As$_2$ layers,
but in addition the magnetic moments of the localized Eu$^{2+}$ order at $T_{\rm N}=19$~K
\cite{Jeevan08,Ren08}. EuFe$_2$As$_2$ has a similar crystallographic and electronic structure
compared to that of SrFe$_2$As$_2$ \cite{Jeevan08}. Therefore, SrFe$_2$As$_2$ can be considered a
non-$f$ homolog of EuFe$_2$As$_2$. In SrFe$_{2-x}$Co$_x$As$_2$ Co-substitution in the Fe$_2$As$_2$
layer stabilizes a superconducting (SC) phase ($0.2\lesssim x\lesssim0.4$) leading to the expectation
that Co-substitution in EuFe$_2$As$_2$ does the same, which in the meantime has been confirmed
experimentally \cite{Jiang09b,Zheng09}.

In this paper we present a pressure study on EuFe$_{1.9}$Co$_{0.1}$As$_2$ by means of
electrical-resistivity ($\rho$) and ac-susceptibility ($\chi_{\rm ac}$) measurements on
polycrystalline samples. In EuFe$_{1.9}$Co$_{0.1}$As$_2$ in our accessible pressure range we can
sensitively tune the magnetic ordering temperature $T_{\rm C}$ from $T_{\rm C}<T_{\rm c,onset}$ to
$T_{\rm C}>T_{\rm c,onset}$ allowing us to study the peculiar interplay of 4{\it f} magnetism and
superconductivity.

\section{Experimental Details}
The polycrystalline samples of EuFe$_{2-x}$Co$_x$As$_2$ were synthesized by sintering stoichiometric
amounts of the precursors EuAs, Fe$_2$As and Co$_2$As. The use of precursors minimizes the elemental
impurities of As and Fe. X-ray diffraction measurements confirmed the BaAl$_4$ type structure (space
group I4/mmm) for all the samples. Electrical resistance and ac susceptibility were measured using an
LR700 resistance/mutual inductance bridge (Linear Research). A miniature compensated coil system
placed inside the pressure cell was utilized for the $\chi_{\rm ac}$ experiments. Temperatures down
to 1.8 K were reached using a physical property measurement system (PPMS, Quantum Design). Pressures
up to 3 GPa have been achieved in a double-layer piston-cylinder type pressure cell with silicone
fluid as pressure transmitting medium. The SC transition of Pb, which served as a pressure gauge,
remained sharp at all pressures indicating good hydrostatic conditions.

\section{Experimental Results}

\begin{figure}[b!]
\begin{minipage}{18pc}
\includegraphics[width=18pc]{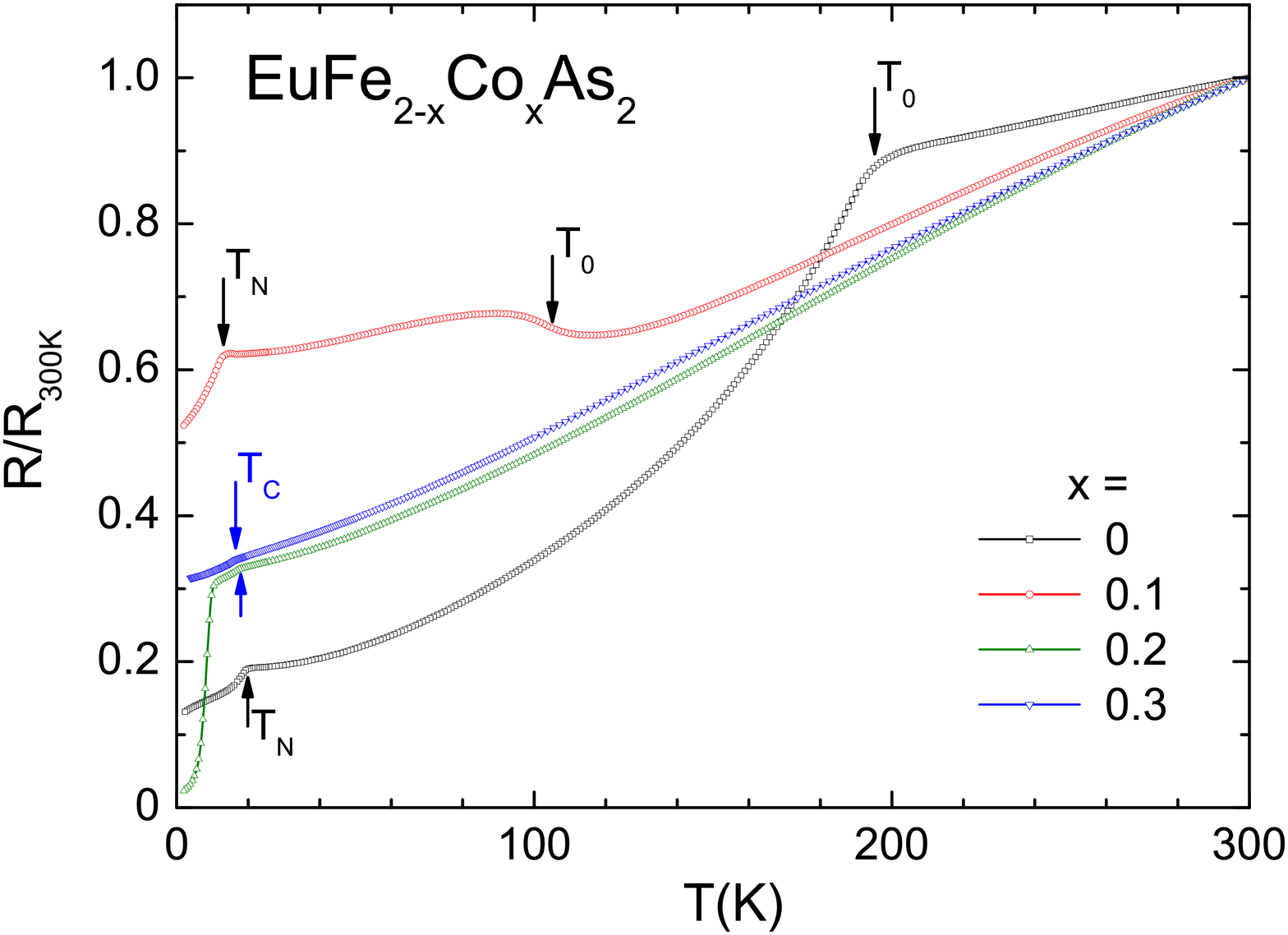}
\caption{\label{rho_x}Electrical resistivity normalized to the value at room temperature ($R/R_{\rm
300K}$) for EuFe$_{2-x}$Co$_x$As$_2$.}
\end{minipage}\hspace{2pc}%
\begin{minipage}{18pc}
\includegraphics[width=18pc]{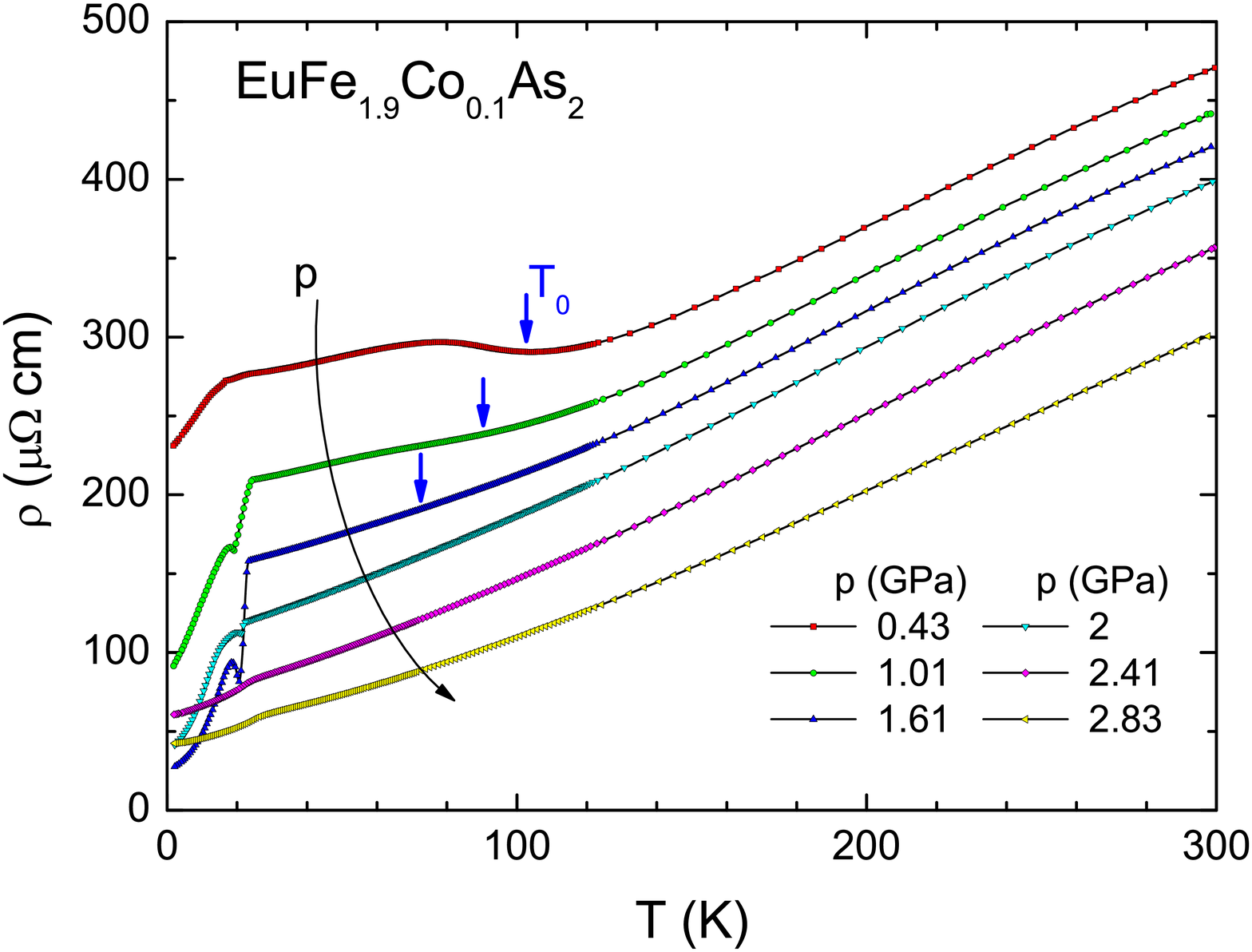}
\caption{\label{rho_p_high}$\rho(T)$ of EuFe$_{1.9}$Co$_{0.1}$As$_2$ at different external
pressures.}
\end{minipage}
\end{figure}

Figure~\ref{rho_x} shows the resistance normalized to the value at room temperature, $R/R_{\rm
300K}(T)$, for EuFe$_{2-x}$Co$_x$As$_2$ ($x=0$, 0.1, 0.2, and 0.3). In EuFe$_2$As$_2$ $R/R_{\rm
300K}(T)$ exhibits two distinct fea\-tures at $T_0=195$~K and $T_{\rm N}=12.5$~K corresponding to the
SDW instability and the ordering of the localized Eu$^{2+}$ moments, respectively, in good agreement
with literature \cite{Jeevan08,Ren08}. Upon Co substitution $T_0$ decreases rapidly to 103~K for
$x=0.1$. At $x=0.2$ no signature of the SDW transition is visible in the $\rho(T)$ data anymore.
$T_{\rm N,C}$ exhibits a weak concentration dependence. In the concentration range between $x\approx
0.1\div0.2$ the type of magnetic order possibly changes from antiferromagnetic (AFM) to ferromagnetic
(FM). In EuFe$_{1.8}$Co$_{0.2}$As$_2$ a sharp drop of the resistance below $\approx$\,10~K indicates
the onset of superconductivity. This observation is consistent with studies on single crystals
\cite{Jiang09b}. However, there a higher $T_c$ and reentrant-like SC behavior was reported. For
$x\leq0.1$ and $x\geq0.3$ no signature of superconductivity is visible in our data evidencing a more
narrow SC region than in the non-$f$ homolog system SrFe$_{2-x}$Co$_x$As$_2$ \cite{Leithe-Jasper08}.

\begin{figure}[t!]
\begin{minipage}{18pc}
\includegraphics[width=18pc]{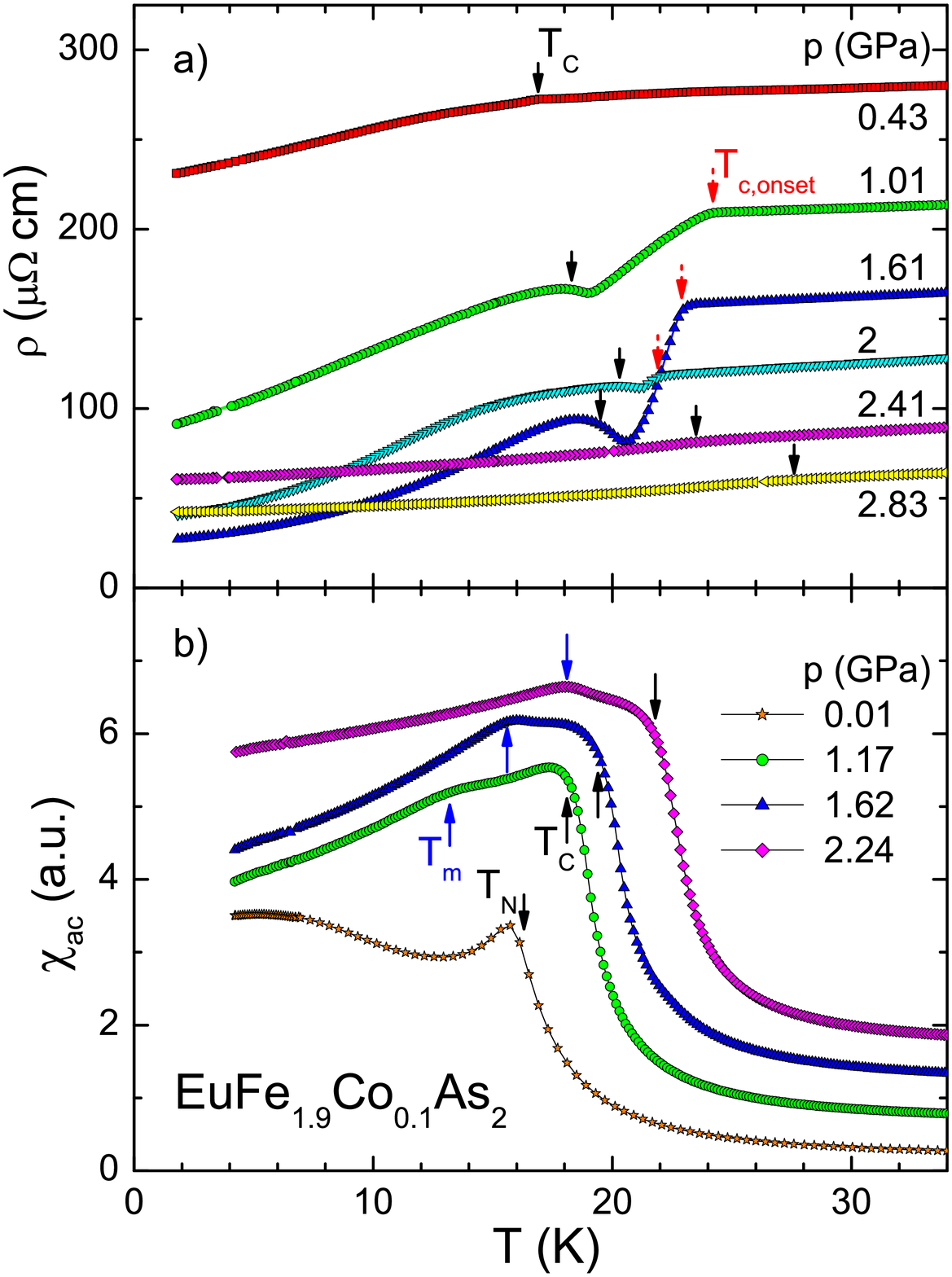}\hspace{2pc}
\end{minipage}\hspace{2pc}%
\begin{minipage}{18pc}
\begin{minipage}{18pc}\caption{\label{rho_chi_p_low}(a) Low-temperature resistivity and (b) ac susceptibility of EuFe$_{1.9}$Co$_{0.1}$As$_2$ at different
external pressures. For clarity the $\chi_{\rm ac}(T)$ data at different pressures have been shifted
with respect to each other.}
\end{minipage}\vspace{2pc}
\includegraphics[width=18pc]{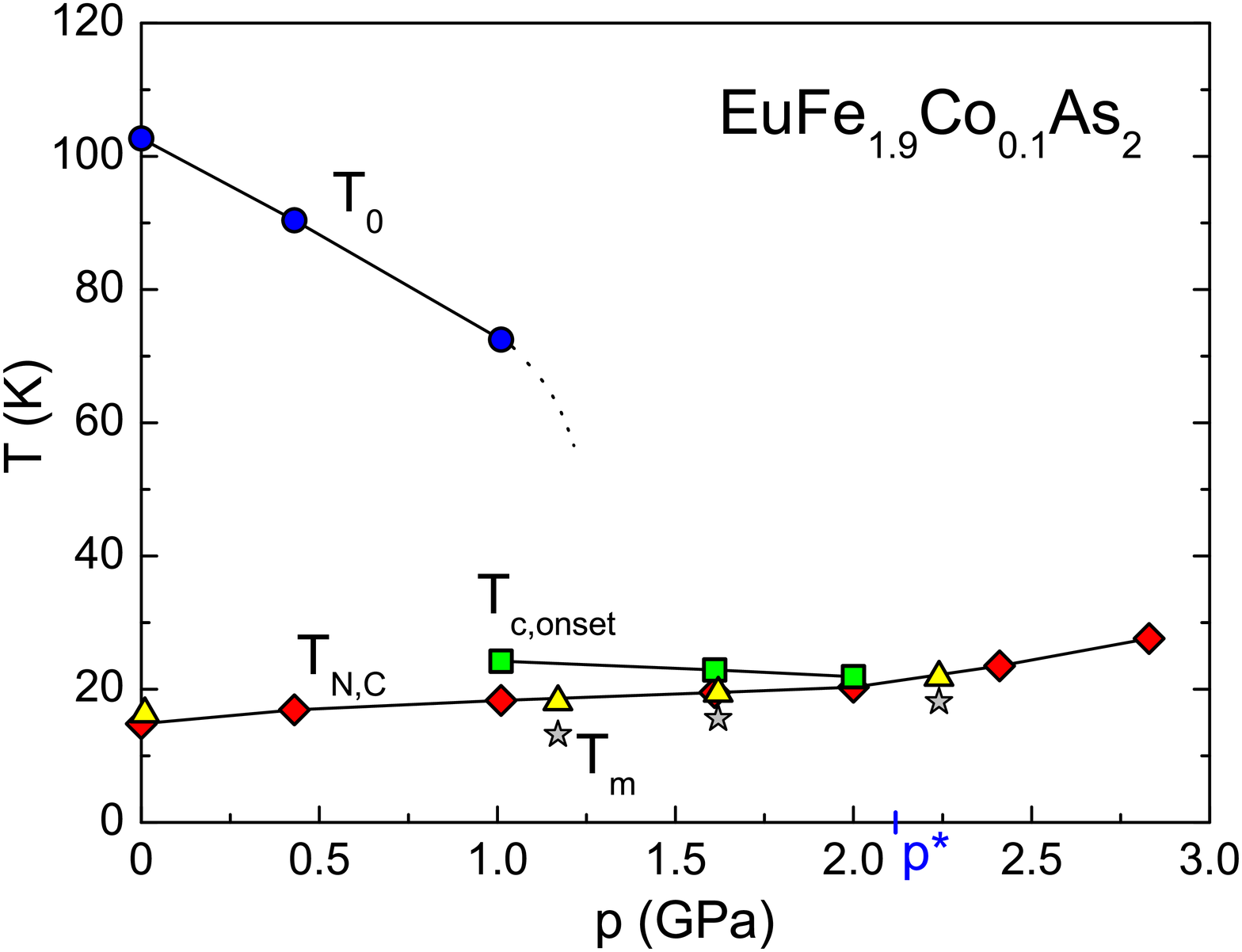}
\caption{\label{PhD}Temperature-pressure phase diagram of EuFe$_{1.9}$Co$_{0.1}$As$_2$.}\vspace{1pc}
\end{minipage}
\end{figure}

In the following, we will focus on the results of our pressure study on EuFe$_{1.9}$Co$_{0.1}$As$_2$.
$\rho(T)$ for different pressures is shown in Fig.~\ref{rho_p_high}. Upon applying pressure the
anomaly indicating the SDW transition at $T_0$ broadens and shifts rapidly to lower temperatures. At
1.61~GPa no feature related to $T_0$ can be resolved in $\rho(T)$ anymore suggesting that the SDW
order is already suppressed at this pressure. At low temperatures, displayed in
Fig.~\ref{rho_chi_p_low}a, a kink in $\rho(T)$ at 14.8~K and 16.9~K indicates the ordering of the
Eu$^{2+}$ moments at 0 and 0.43 GPa, respectively \cite{TNTC}. At 1.01~GPa a first signature of a SC
transition becomes evident: $\rho(T)$ drops sharply below $T_{\rm c,onset}=24.2$~K followed by a
small maximum indicating the magnetic ordering of the Eu$^{2+}$ and a further decrease in $\rho(T)$
on lowering the temperature. This behavior is similar to that observed previously, but at higher
pressures in EuFe$_2$As$_2$ \cite{Miclea09}. The drop in $\rho(T)$ as well as the following maximum
are much more pronounced at $p=1.61$~GPa, but are hardly visible anymore at $p=2$~GPa. At even higher
pressures no indication of superconductivity is present in the resistivity data anymore.

The results of the $\chi_{\rm ac}(T)$ experiments are presented in Fig.~\ref{rho_chi_p_low}b. The
$\chi_{\rm ac}(T)$ curve at the lowest pressure ($p=0.01$~GPa) and the curves at higher pressures are
qualitatively different. While the shape of $\chi_{\rm ac}(T)$ at $p=0.01$~GPa is reminiscent of an
AFM phase transition, the shape at $p>0.01$~GPa is typical for a FM transition. A change of the
magnetic groundstate from AFM to FM has been found in EuFe$_2$Co$_2$As$_2$ on P-doping on the As-site
too \cite{Ren09}. P doping corresponds to the application of positive chemical pressure, therefore,
our finding is not surprising and in agreement with the doping studies. On increasing pressure the
anomaly in $\chi_{\rm ac}(T)$ at $T_{\rm N,C}$ shifts to higher temperatures \cite{TNTC}. At
pressures $p>0.01$~GPa a second feature appears in $\chi_{\rm ac}(T)$  below $T_{\rm C}$ at $T_{\rm
m}$ \cite{TNTC}. We attribute this kink to a change in the magnetic structure from FM to AFM. No
feature in $\rho(T)$ is visible at $T_{\rm m}$. A detailed study of the magnetic properties will be
presented elsewhere. It is important to note that we do not observe a clear signature of
superconductivity in our $\chi_{\rm ac}$ results.

In our study we observe a less robust superconductivity than reported in literature
\cite{Jiang09b,Zheng09}. This might be related to different sample preparation procedures or to
internal strain in the samples. An important influence of internal/external residual strain on the
observed physical properties appears also in our experiments. After releasing the pressure, we still
find reentrant SC behavior in $\rho(T)$ similar to the data at 1.61~GPa (not shown).

The $T-p$ phase diagram in Fig.~\ref{PhD} summarizes our results. While $T_{\rm c}(p)$ decreases only
slightly with increasing pressure $T_{\rm N,C}(p)$ increases strongly by a factor of nearly 2 from
14.8~K at ambient pressure to 27.6~K at 2.83~GPa, the highest pressure of our experiment. We find a
good agreement between $T_{\rm N,C}(p)$ obtained from the $\rho(T)$ and $\chi_{\rm ac}(T)$
experiments. The strong increase of $T_{\rm N,C}(p)$ with pressure has not been reported in
EuFe$_2$As$_2$ \cite{Miclea09}. There, the magnetic ordering temperature is nearly independent of
pressure. The different pressure dependencies of $T_{\rm c,onset}(p)$ and $T_{\rm N,C}(p)$ in
EuFe$_{1.9}$Co$_{0.1}$As$_2$ lead to a crossing of both temperature lines [$T_{\rm c,onset}(p)=T_{\rm
C}(p)$] in the $T-p$ phase diagram at a pressure $p^*$ slightly higher than 2~GPa. Above $p^*$ once
$T_{\rm C}>T_{\rm c}$ no indication of any SC transition is visible in the resistivity data. Already
at 2~GPa where $T_{\rm c}=21.9$~K and $T_{\rm C}=20.3$~K get rather close. The feature associated to
the SC transition is considerably reduced compared to the previous pressure.

\section{Discussion and Summary}

In summary our results show that in the substitution series EuFe$_{2-x}$Co$_x$As$_2$ the SDW
transition temperature $T_0$ is suppressed rapidly upon increasing Co concentration, while the
ordering temperature of the Eu$^{2+}$ moments is almost constant. For $x=0.2$ we observe the onset of
superconductivity in $\rho(T)$ at $\approx$\,10~K. For $x<0.2$ and $x>0.2$ no indication of
superconductivity is present. The strong suppression of $T_0$ as function of the Co concentration and
the appearance of superconductivity in the $T-x$ phase diagram is consistent with the expectation
from the comparison with the homolog non-$f$ Co-substitution series SrFe$_{2-x}$Co$_x$As$_2$
\cite{Leithe-Jasper08}. However, there superconductivity is stable in a broader concentration range.
The presence of only a narrow SC regime in the Eu system might be related to the presence of the
local moment Eu$^{2+}$ magnetism.

Pressure rapidly suppresses the SDW instability at $T_0$ in EuFe$_{1.9}$Co$_{0.1}$As$_2$. In the same
pressure region where the signature of $T_0$ is lost in $\rho(T)$ the onset of superconductivity is
observed. At low temperatures our results suggest a strong detrimental effect of the Eu$^{2+}$
magnetism on the superconductivity. In an intermediate pressure range ($1.01~{\rm GPa}\leq
T\leq2~{\rm GPa}$) a the system is on the verge to a SC state, indicated by the drop in the
resistivity below $T_{\rm c,onset}$. However, the formation of the SC state is inhibited by the
formation of long range magnetic order of the Eu$^{2+}$ moments. A similar result has been reported
previously in undoped EuFe$_2$As$_2$ \cite{Miclea09}. The important difference in the present study
on EuFe$_{1.9}$Co$_{0.1}$As$_2$ is that upon increasing pressure $T_{\rm N,C}(p)$ increases while
$T_{\rm c,onset}(p)$ decreases and, therefore, $T_{\rm C}(p)$ becomes greater than $T_{\rm
c,onset}(p)$ above a certain pressure $p^*$. Once $T_{\rm C}(p)>T_{\rm c,onset}(p)$ no signature of
superconductivity is found anymore. We take this as an evidence for the strong detrimental effect of
the Eu$^{2+}$ magnetism on the the formation of the SC state. As our ac-susceptibility results
indicate Eu$^{2+}$ orders ferromagnetically under pressure. Thus, we speculate that the internal
magnetic fields due to the ferromagnetically ordered Eu$^{2+}$ moments suppress the superconductivity
in the iron-arsenide layers.

\section*{Acknowledgements}

We gratefully acknowledge the Deutsche Forschungsgemeinschaft (DFG) for financial support through SPP
1458.


\begin{thebibliography}{99}
\bibitem{TNTC} $T_{\rm N,C}$ and $T_{c, {\rm onset}}$ were defined from $\rho(T)$ by taking the position of the corresponding
minimum in ${\rm d}^2\rho(T)/{\rm d}T^2$. $T_{\rm N,C}$ and $T_{\rm m}$ from $\chi_{\rm ac}(T)$ were
determined accordingly.
\end{thebibliography}
\end{document}